\documentclass[aps,prb,groupeaddress,twocolumn]{revtex4-1}
\usepackage{graphicx}
\begin{document}

\title{Atomic configuration and properties of austenitic steels at
finite temperature: The effect of longitudinal spin fluctuations}

\author{A. V. Ruban}
\affiliation{Department of Materials Science and Engineering, KTH Royal
Institute of Technology, SE-100 44 Stockholm, Sweden}
\affiliation{Materials Center Leoben Forschung GmbH, A-8700 Leoben,
Austria}

\author{M. Dehghani}
\affiliation{Materials Center Leoben Forschung GmbH, A-8700 Leoben,
Austria}

\date{\today}

\begin{abstract}
High temperature atomic configurations of fcc Fe-Cr-Ni alloys with
alloy composition close to austenitic steel are studied in statistical
thermodynamic simulations with effective interactions obtained in
{\it ab initio} calculations.
The latter are done taking longitudinal spin fluctuations (LSF) into
consideration within a quasiclassical phenomenological model. It is
demonstrated that magnetic state affects greatly the  alloy properties
and in particular, it is shown that the LSF substantially modify the
bonding and interatomic interactions of fcc Fe-Cr-Ni alloys even at
ambient conditions. The calculated atomic short-range order (SRO) is
in reasonable agreement with existing experimental data for
Fe$_{0.56}$Cr$_{0.21}$Ni$_{0.23}$, which has strong preference for
the (001) type ordering between Ni and Cr atoms. A similar ordering
tendency is found for the Fe$_{0.75}$Cr$_{0.17}$Ni$_{0.08}$ alloy
composition, which approximately corresponds to the widely used
304 and 316 austenitic steel grades.
\end{abstract}


\maketitle

\section{Introduction}

Austenitic steels based on Fe-Cr-Ni system are in extensive use
in different domestic and industrial applications due to their
excellent corrosion resistant and mechanical properties. 
They usually consist of more than three elements added on a
purpose or accidently, however, Cr and Ni are the usual alloying
elements: Cr provides the corrosion resistance, and
its content is usually in between 15--25 at. \%, while Ni stabilizes
the fcc structure, and its content is as a rule within 5-15 at.\%
(can be up to 35 at.\% in some special grades).
Although these alloys have been for decades under development
and investigation, accurate description of their finite 
temperature properties at the atomic and electronic structure levels
is still a challenging task.

One of the main obstacles in getting accurate {\em ab initio}
picture of fcc Fe-Cr-Ni alloys is their non-trivial magnetism.
The magnetic phase diagram of Fe$_{80-x}$Ni$_x$Cr$_{20}$
($10 \leq x \leq 30$) alloys have been determined in
Ref. \onlinecite{mag_phd}. At low temperatures, one can find
antiferromagnetic, spin-glass, ferromagnetic and a mixture of
ferromagnetic and spin glass states for a certain range of compositions.
In particular, the antiferromagnetic state is the low temperature
magnetic state with the Neel temperature of about 20-40 K if
concentration of Ni is within  5-15 at. \%. This means that in
practice austenitic steels are in paramagnetic state at ambient
conditions.

The problem is that this paramagnetic state is a highly non-trivial
phenomenon, which is extremely difficult to model accurately on
the first-principles level since a static disordered local moment
model breaks down due to disappearance of the local magnetic moments
on Cr and Ni atoms in the density functional theory (DFT) calculations.
At the same time, their magnitude deviates substantially from zero
due to longitudinal spin fluctuations on short time scales,
producing a significant effect upon all the properties of steels. 
In principle, such magnetic excitations can be accounted
for using advanced {\it ab initio} methods like, for instance,
dynamical mean-field theory.\cite{dmft} However, their applications
to steels, which are multicomponent random alloys, is too cumbersome.

In this paper, we therefore use a simplified formalism for the LSF
developed in Ref. \onlinecite{ruban07_lsf,ruban13}.
It is a classical high temperature limit of the spin-fluctuation
theory\cite{moriya78,moriya85} where the contribution due to
thermally induced spin fluctuations are considered within a
phenomenological model based on a classical magnetic Hamiltonian.
Although it breaks down at low temperatures where quantum effects
are important, it provides a reasonable account of the 
LSF at elevated temperatures.\cite{ruban13}

Using this model of the LSF, we consider properties of Fe-Cr-Ni alloys
and, in particular, atomic ordering at elevated temperatures. The
experimental information on the atomic short range order (SRO) in
austenitic steels is scarce. A quite detailed investigation
of the atomic SRO in fcc Fe$_{56}$Cr$_{21}$Ni$_{23}$
was done by Cenedese {\em et al.}\cite{cenedese84} by thermal neutron
diffuse scattering from single crystals. The atomic SRO has been also
measured in alloys with similar compositions, Fe$_{82-x}$Cr$_{18}$Ni$_{x}$
($x=$ 15, 20, and 25 wt.\%) by Braude {\it et al.} using
x-ray diffuse scattering technique.\cite{braude86}  However, some 
results of this investigation are contradictory as is discussed
below.

Recently, the phase stability of ternary fcc and bcc Fe-Cr-Ni alloys
has been investigated in Ref. \onlinecite{wrobel15} using a
combination of DFT, cluster expansion (CE),
and magnetic cluster expansion (MCE) techniques. These authors have
found good agreement of different calculated properties with experimental
data, including results for the atomic SRO in Fe$_{56}$Cr$_{21}$Ni$_{23}$
alloy as stated by the authors. At the same time, the results presented
in Table VII of Ref. \onlinecite{wrobel15} show some obvious
problems for the calculated atomic SRO in this alloy. Namely, the
calculated atomic SRO is too strong compared with the
experimental data,\cite{cenedese84} especially taking into
consideration the fact that calculations are for 1300 K, while
in the experiment, the samples were annealed first for 1 hour at
1273 K and then for 10 hours at 773 K. Of course, it might well be
that 10 hours is not enough to fully equilibrate this alloy at 773 K,
nevertheless a substantial rearrangement can be expected on a local
scale of at least several interatomic distances. In fact, the
authors of Ref. \onlinecite{wrobel15} find ordering transition
at 1550 K for Fe$_{50}$Cr$_{25}$Ni$_{25}$. At this temperature
the diffusion is very fast, and had such a high-temperature
transition really existed, it or the corresponding ordered phase
would be definitely seen or detected in different kind of
experiments.

This means that there is a problem with the existing theoretical
description of the atomic SRO in austenitic steels. To solve it is
one of the aims of the present investigation. Another aim is to
demonstrate the role of the LSF in austentic steels at finite
temperatures. For that purpose, we consider here two alloy:
Fe$_{75}$Cr$_{17}$Ni$_{08}$ alloy, whose composition
is close to the widely used 304 and 316 steel grades,
and Fe$_{56}$Cr$_{21}$Ni$_{23}$, whose atomic SRO has been
obtained experimentally\cite{cenedese84} and just
recently calculated using {\it ab initio} theory.\cite{wrobel15}
We also outline a technique for calculating chemical and magnetic
exchange interactions  within the exact muffin-tin orbital (EMTO)
method.\cite{andersen94,vitos07}

\section{First-principles methodology}

Electronic structure calculations of random fcc Fe-Cr-Ni alloys have
been done using the coherent potential approximation (CPA)\cite{CPA}
and locally self-consistent Green's function (LSGF) technique,\cite{lsgf}
which accurately accounts for the local environment effects in random
alloys.
Both these techniques have been used within the EMTO
method\cite{andersen94,vitos07} referenced here as
EMTO-CPA\cite{vitos01} and ELSGF\cite{peil12}, respectively.
The EMTO-CPA calculations have been done by the Lyngby version
of the Green's function EMTO code,\cite{code} where the screened
Coulomb interactions in the single-site DFT-CPA
approximation\cite{screening} and screened generalized perturbation
method (SGPM)\cite{ducastelle76,ducastelle,ruban04} are implemented
(See Appendix \ref{sgpm}). 

In particular, the contributions of the screened Coulomb interactions
in the DFT-CPA to the one-electron potential of alloy
components, $V^i_{\rm scr}$, and to the total energy, $E_{\rm scr}$,
are:\cite{screening}

\begin{eqnarray}
V^i_{\rm scr} &=& -e^2 \alpha^0_i\frac{\bar{q}_i}{S}\nonumber \\
 E_{\rm scr} = \sum_i c_i E^i_{\rm scr};  \quad
 E^i_{\rm scr} &=& -e^2 \frac{1}{2}\alpha^0_i
\beta_{\rm scr}\frac{\bar{q}^2_i}{S}.
\end{eqnarray}
where $\bar{q}_i$ and $\alpha_i^0$ are the net charge of the atomic
sphere of the $i$th alloy component in the single-site CPA
calculations and its on-site screening constant, which are different
for different alloy components in multicomponent alloys, $S$ the
Wigner-Seitz radius, $\beta_{\rm scr}$ the average on-site screening
constants, which accounts for the electrostatic multipole moment energy
contribution due to inhomogeneous local environment of different sites
in random alloy.

The self-consistent electronic structure calculations have
been done within the local density approximation using Perdew
and Wang functional,\cite{LDA92} while the total energy have 
been calculated using the full charged density
technique\cite{vitos07} in the PBE generalized gradient
approximation.\cite{pbe96} In the Brillouin zone integration, a
32$\times$32$\times$32 Monkhorst-Pack grid have been
used.\cite{monkhorst72}
All the calculations have been done with $l_{max}=3$ for
partial waves and the electronic core states were recalculated
at every iteration during the self-consistent calculations for
valence electrons.

The on-site and inter-site screening constants needed in the
EMTO-CPA and SGPM calculations were obtained for 
Fe$_{75}$Cr$_{17}$Ni$_{08}$ and Fe$_{50}$Cr$_{25}$Ni$_{25}$
random alloys in a 864-atom supercell (6$\times$6$\times$6
translations of the fcc 4-atom cubic unit cell) ELSGF calculations.
After the configurational optimization, the first six Warren-Cowley
SRO parameters of the supercells were less than 0.002 (in absolute value)
and next 2, for the 7th and 8th coordination shells, less than 0.015
for all three pairs of the alloy components. 
The calculations were done in the disordered local
moment state\cite{cyrot72,gyorffy85} as is implemented in the ELSGF
code\cite{peil12} with the LSF at 800 K (see the next section).
The local interaction zone (LIZ) has contained
the first two coordination shells of the central site (of the LIZ).

The on-site screening constants have been determined as

\begin{equation} \label{a_0_1}
\alpha^0_i = \frac{S \langle V^{\rm Mad}_i \rangle}
{e^2 \langle q_i \rangle} ,
\end{equation}
where $\langle q_i \rangle$ and $\langle V^{\rm Mad}_i \rangle$
are the conditional average of the net charges, $q_i$,
and the Madelung potentials, $V^{\rm Mad}_i$ of the
$i$-th component in the supercell.
The calculated on-site screening constants, $\alpha^0_i$, vary very
little with alloy composition, lattice constant, and temperature
due to thermal electronic and magnetic excitations due to the LSF. They
are approximately equal to 0.725, 0.777, and 0.823 for Fe, Cr, and
Ni, respectively, while $\beta_{\rm scr}$ is about 1.14. 
These values of the screening constants have been used
in all the DFT-CPA calculations.

The intersite screening constants, $\alpha^{ij}_{p}$,
needed in the calculations of the electrostatic contribution,
$V^{ij-\rm scr}_{p}$ to the SGPM potential at the $p$th coordination
shell for the  $i$-$j$ pair of alloy components:

\begin{equation} \label{eq:V_scr}
V^{ij-\rm scr}_{p} = e^2\alpha^{ij}_{p}\frac{\bar{q}^2_{ij}}{S} ,
\end{equation}
where $\bar{q}_{ij}=\bar{q}_i - \bar{q}_j$.
Screening constants $\alpha^{ij}_{p}$
have been obtained in the supercell ELSGF calculations 
random alloy 
from the screening charge by exchanging the
corresponding alloy components ($i$ and $j$ in some specific
sites having random local environment on average) as is described
in Ref. \onlinecite{screening}.

\section{Longitudinal spin fluctuations in
F\lowercase{e}-C\lowercase{r}-N\lowercase{i} alloys}

The description of the paramagnetic state in fcc alloys in
DFT calculations at ambient conditions requires an additional
modeling, which takes into consideration thermally induced
longitudinal spin
fluctuations.\cite{staunton92,uhl96,rosengaard97,kubler00,ruban07_lsf,ruban13} 
Here, we follow the formalism developed in
Refs. \onlinecite{ruban07_lsf} and  \onlinecite{ruban13}. 
The main idea is to consider the LSF as an entropic effect
within a static one-electron consideration.
In this case, the magnitude of the local magnetic 
moment induced by the entropy can be obtained, to a first
approximation, from the longitudinal spin fluctuation energy
in the corresponding statistical thermodynamic simulations.

The longitudinal spin fluctuation energy is the energy of
embedding an impurity having a given magnetic moment
$m_i$ into the DLM effective medium with a certain choice of
the average local magnetic moments of alloy components.\cite{ruban07_lsf}
Since, the average local magnetic moment at a given
temperature is not known in advance, the statistical
simulations have to be done in a self-consistent way.

Obviously, this is an extremely time consuming scheme in
the case of multicomponent alloys. However, it can be 
greatly simplified by using an approximate expression for
the entropy of longitudinal spin fluctuations:\cite{ruban13}

\begin{equation} \label{eq:S_lsf}
S_i^{\rm lsf} = 3 \ln m_i ,
\end{equation}
which is valid in the classical high-temperature limit for
the quadratic form of the LSF energy. Although the later is
approximately true only for Ni and Cr, while for Fe there is a
non zero local magnetic moment for equilibrium lattice constants
of austenitic steels, we have used this expression for Fe too
in order to keep continuous description of the magnetic
energy for small lattice constants.

\begin{figure}
\includegraphics[width=0.9\linewidth]{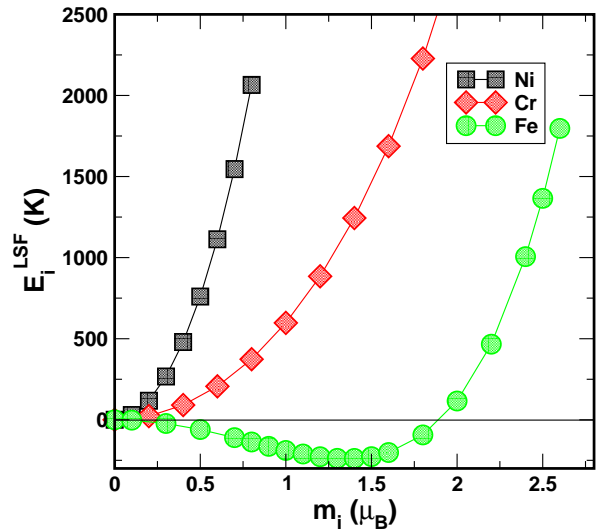}
\caption{\label{fig:E_lsf} (Color online) 
Longitudinal spin fluctuation energies of alloy components
in Fe$_{75}$Cr$_{17}$Ni$_{08}$ in the DLM state obtained
for the Fe, Cr, and Ni local magnetic moments 1.69, 0.82, and
0.38 $\mu_{\rm B}$, respectively.}
\end{figure}

In Fig. \ref{fig:E_lsf}, the longitudinal spin fluctuation 
energy for Fe, Cr, and Ni in Fe$_{75}$Cr$_{17}$Ni$_{08}$ in
the DLM effective medium due to the LSF is shown. In these calculations,
the local magnetic moment of one of the components have been
changed while all the other were kept fixed to the following
magnitudes: 1.69 $\mu_{\rm B}$ for Fe, 0.82 $\mu_{\rm B}$ for
Cr, and 0.38 $\mu_{\rm B}$ for Ni, which approximately corresponds
to their local magnetic moments at 300 K obtained using Eq.
(\ref{eq:S_lsf}) in the self-consistent calculations.
One can see, that the LSF energy curves for Ni and Cr resemble parabola.
In the case of Fe, the LSF energy has minimum at $\approx$ 1.4 $\mu_B$.
Thus, at least a fourth order polynomial is needed to get a
qualitative behavior of the LSF energy in the latter case. 

\begin{figure}
\includegraphics[width=0.9\linewidth]{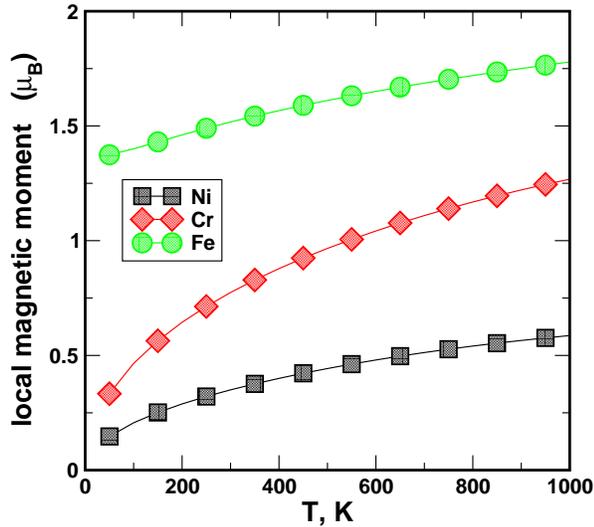}
\caption{\label{fig:m_lsf} (Color online) 
Local magnetic moments of alloy components in
in Fe$_{0.75}$Cr$_{0.17}$Ni$_{0.08}$ due to LSF obtained from
the LSF energies presented in Fig. \ref{fig:E_lsf}
as a function of temperature.}
\end{figure}

However, this approximate scheme works reasonably well even for
the LSF induced magnetic moment of Fe. In Fig. \ref{fig:m_lsf},
the local magnetic moments of Ni, Cr, and Fe are shown as function
of temperature, which have been obtained in the single-site 
mean-field approximation from the LSF energies, $E_i^{\rm LSF}$,
presented in Fig. \ref{fig:E_lsf} as:

\begin{equation} \label{eq:m_lsf}
m_i = 1/Z_i \int E_i^{\rm LSF}(m) m^3 dm ,
\end{equation}
where $Z_i$ is the partial statistical sum for alloy 
component $i$:

\begin{equation} \label{eq:m_lsf}
Z_i = \int E_i^{\rm LSF}(m) m^2 dm ,
\end{equation}

It can be substantially simplified without losing much accuracy
(which is not on the demand in semiquantitative modeling) by
just calculating the energy of the DLM state with varying magnitude
of the local magnetic moment of one of the alloy components while
keeping m of the others fixed to the one which corresponds to
the chosen temperature. For instance, using (\ref{eq:S_lsf})
one finds that the
local magnetic moments in Fe$_{75}$Cr$_{17}$Ni$_{08}$ at 300 K
are 1.69, 0.82, and 0.38 $\mu_B$ for Fe, Cr, and Ni, respectively,
while they are 1.52, 0.77, and 0.35
$\mu_B$ if (\ref{eq:m_lsf}) is used with the corresponding 
spin-fluctuation energies. As one can see, even for Fe,
Eq. (\ref{eq:S_lsf}) produces quite reasonable results only
slightly overestimating local magnetic moment. Let us note that
without LSF the local magnetic moment of Fe is about 1.4 $\mu_B$
while Cr and Ni becomes non-spin polarised. 

\section{Lattice parameter and elastic constants of 
 F\lowercase{e}$_{75}$C\lowercase{r}$_{17}$N\lowercase{i}$_{08}$
 at ambient conditions}

\begin{figure}
\includegraphics[width=0.9\linewidth]{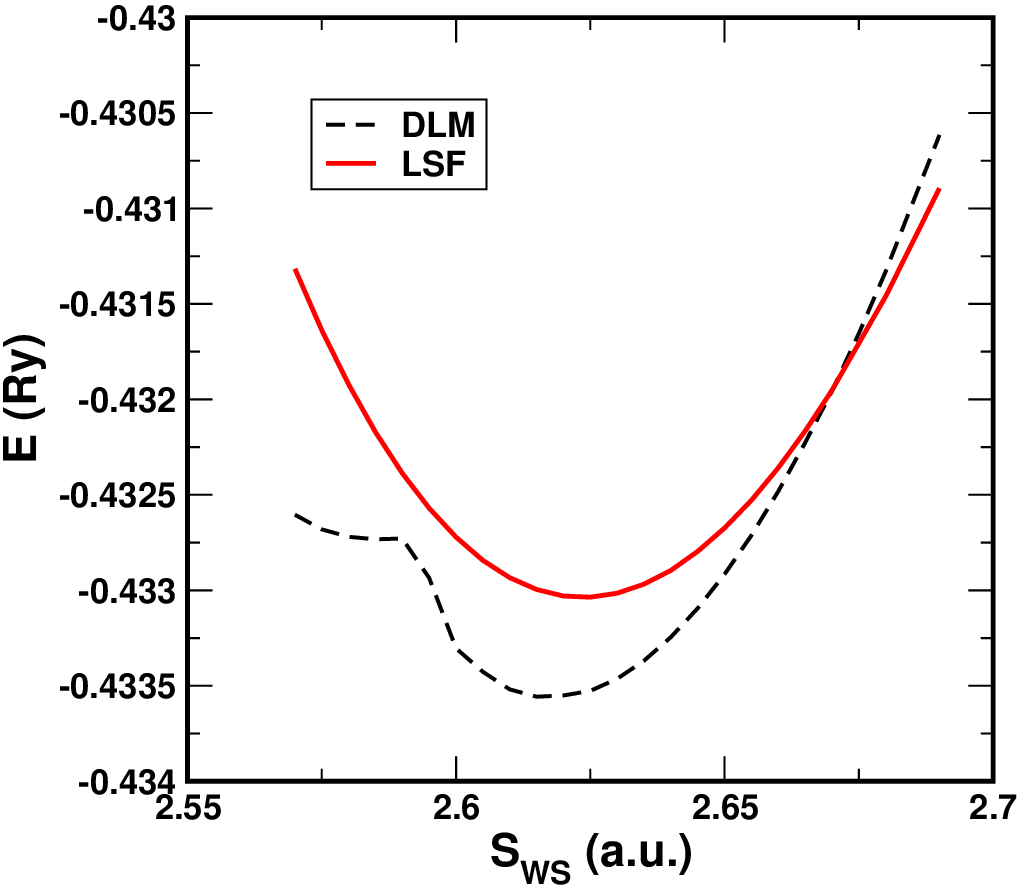}
\caption{\label{fig:E_tot} (Color online) 
The electronic and magnetic free energies of
Fe$_{75}$Cr$_{17}$Ni$_{08}$ at 300 K obtained
in the DLM calculations with and without LSF. The free energies
are shifted to some arbitrary constant.}
\end{figure}

The importance of the LSF in austenitic steels even at room
temperature can be seen in the first-principles calculations
of the lattice constant and bulk modulus of Fe$_{75}$Cr$_{17}$Ni$_{08}$ 
alloy. In Fig. \ref{fig:E_tot}, we show the total energies 
(up to an arbitrary constant) of random Fe$_{75}$Cr$_{17}$Ni$_{08}$ 
alloy obtained in the EMTO-CPA calculations in the DLM 
paramagnetic state with and without LSF. In the DLM calculations,
the magnetic entropy of Fe has been accounted for using the
entropy of the ideal paramagnetic gas
($S_{mag}^{\rm IP} = \ln(m_{Fe} + 1)$). The one-electron thermal
excitations have been included using Fermi-Dirac distribution
function.\cite{mermin65}

\begin{figure}
\includegraphics[width=0.9\linewidth]{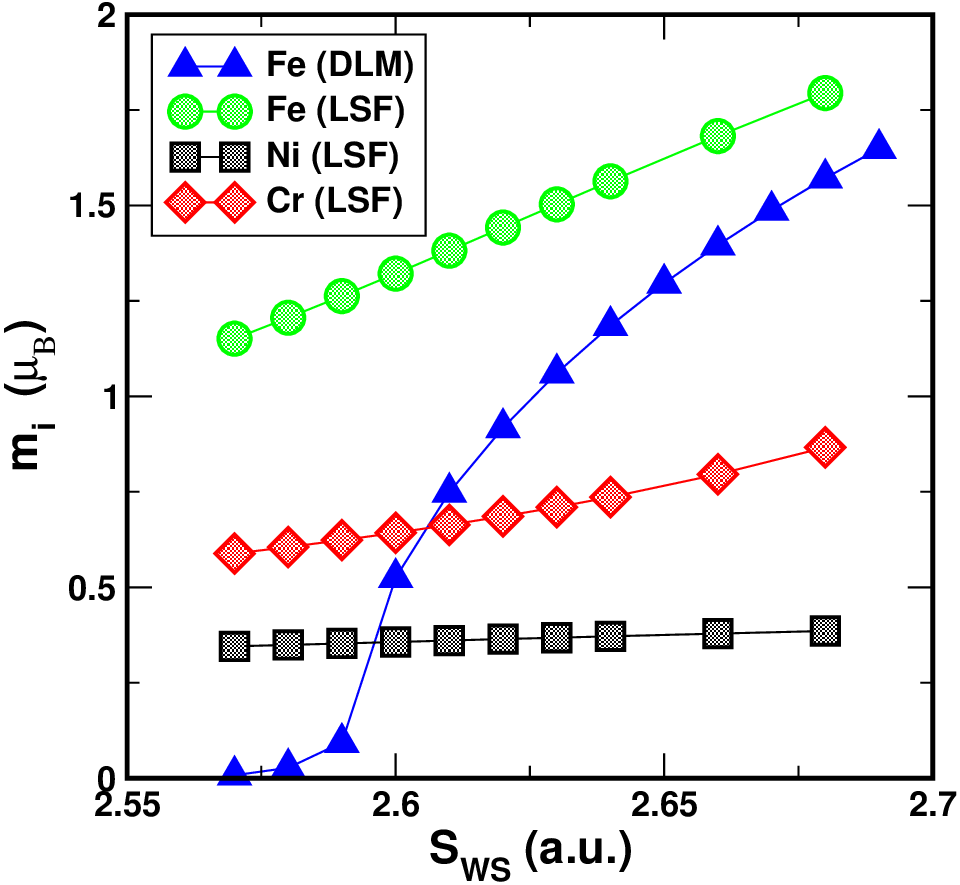}
\caption{\label{fig:m_S} (Color online) 
Local magnetic moments of Fe, Cr, and Ni in
Fe$_{75}$Cr$_{17}$Ni$_{08}$ in the DLM calculations
with and without LSF.}
\end{figure}

If LSF are not included, the free energy curve exhibits quite
irregular behaviour around Wigner-Seitz (WS) radius of 2.6 a.u. 
(or lattice constant about 3.52 \AA), which can be traced
down to the abrupt change of the magnitude of magnetic moment
of Fe, which is clearly seen in Fig. \ref{fig:m_S}. 
The equilibrium WS sphere radii (without phonon contribution)
is 2.618 a.u. and bulk modulus is about 210 gPa, which
is too large compared to the room temperature bulk modulus of
austenitic steels with similar composition, which is usually
in the range of  140-170 GPa.\cite{teklu04} 

The inclusion of the LSF, apart from smoothing the total energy
curve, leads to the increase of the equilibrium WS radius to
2.623 a.u. or lattice constant of 3.55 \AA ~and to a substantial
decrease of the bulk modulus:
161 GPa. In the Debye-Gr\"uneisen model, the room temperature
lattice constant then comes out to be $~$3.57 \AA ~and bulk
modulus 155 GPa, which are in reasonable agreement with
experimental data.\cite{teklu04}  The calculated room temperature
shear elastic constants, $c'$ and $c_{44}$: 35 and 138 GPa, 
respectively, are also in
good agreement with the existing experimental data: 38 and
121 GPa.\cite{teklu04}

\section{Effective interactions}

Finite temperature atomic configuration of Fe$_{75}$Cr$_{17}$Ni$_{08}$
and Fe$_{56}$Cr$_{21}$Ni$_{23}$ has been obtained in Monte
Carlo simulations using the following configurational Hamiltonian:

\begin{eqnarray} \label{eq:H_conf}
H &=& \frac{1}{2} \sum_p \sum_{\alpha, \beta \neq \delta}
V^{(2)-\alpha \beta [\delta]}_p 
\sum_{ij \in p} \delta c^{\alpha}_i \delta c^{\beta}_j + \\ \nonumber
&&\frac{1}{3} \sum_t \sum_{\alpha, \beta, \gamma \neq \delta}
V^{(3)-\alpha \beta \gamma [\delta]}_t
\sum_{i,j,k} \delta c^{\alpha}_i \delta c^{\beta}_j \delta c^{\gamma}_k + h.o.t. .
\end{eqnarray}
Here, the summation is performed over alloy different type of 
clusters ($p$ and $t$ stands for indexes of the pairs and triangles),
alloy components (designated by Greek letters) and lattice sites ($i$,
$j$, and $k$);  $V^{(2)-\alpha \beta [\delta]}_p$ and 
$V^{(3)-\alpha \beta \gamma [\delta]}_t$ are the pair- and 
three-site effective interactions, which have been determined using the
SGPM implemented in the Lyngby version of the EMTO-CPA code (see Appendix
for details), and
$\delta c^{\alpha}_i = c^{\alpha}_i - c^{\alpha}$ is the 
concentration fluctuation of the $\alpha$ component from its
average concentration in alloy, $c^{\alpha}$ at site $i$.

The contribution from pair interactions in (\ref{eq:H_conf}) can 
be reduced to a quasibinary form:

\begin{equation} \label{eq:H_conf_qb}
H^{(2)} = -\frac{1}{2} \sum_p \sum_{\alpha \neq \beta} 
\widetilde{V}^{(2)-\alpha \beta}_p 
\sum_{ij \in p} \delta c^{\alpha}_i \delta c^{\beta}_j ,
\end{equation}
where $\widetilde{V}^{(2)-\alpha \beta}_p$ are the usual binary
effective interactions describing the mutual ordering of $\alpha$
and $\beta$ atoms and connected with the multipcomponent effective
pair interactions, $V^{(2)-\alpha \beta [\delta]}_p$
as\cite{ruban97,V_quasib}

\begin{equation}
V^{(2)-\alpha \beta [\delta]}_p = \frac{1}{2} \left[ 
\widetilde{V}^{(2)-\alpha \delta}_p +
\widetilde{V}^{(2)-\beta  \delta}_p -
\widetilde{V}^{(2)-\alpha \beta}_p \right] .
\end{equation}

The advantage of such a quasibinary representation is
its direct connection to the Hamiltonians and interactions of
the binary systems of the components composing the multicomponent
one. Unfortunately, the contribution from multisite interactions
cannot be reduced to a similar quasibinary form due to the presence
of additional indexes of alloy components. 

In first-principles calculations, the interactions entering Eqs.
(\ref{eq:H_conf}) and (\ref{eq:H_conf_qb}) have been obtained 
using a DLM six component model of the three-component alloys: 
two components with the opposite orientation of magnetic moment
for each alloy component. The chemical interactions have been
obtained in the magnetic-moment averaged form for each alloy
component in the DLM-CPA calculations. The corresponding
expressions for the interactions in the EMTO-CPA method are
given in Appendix \ref{sgpm}.

The effective interactions of the above Hamiltonian are not only
concentration dependent, but they also depend on the temperature
due to the temperature dependence of the equilibrium volume of 
the alloy, its magnetic state, and the local magnetic moments of
its components in the DLM-LSF state. As is demonstrated below such
a temperature dependence should be taking into consideration in
atomistic modeling at high temperatures.

\begin{figure}
\includegraphics[width=0.9\linewidth]{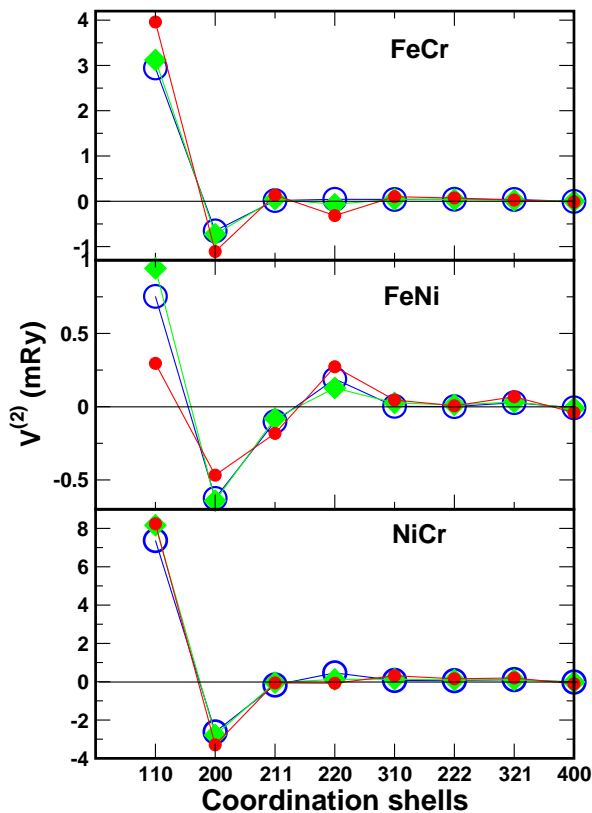}
\caption{\label{fig:V2} (Color online) 
Quasibinary effective pair interactions in 
Fe$_{75}$Cr$_{17}$Ni$_{08}$ (open circles) and
Fe$_{56}$Cr$_{21}$Ni$_{23}$ (filled diamonds) obtained at
800 K in the DLM-LSF calculations.
Small filled circles show the results for Fe$_{56}$Cr$_{21}$Ni$_{23}$
without LSF (which means DLM calculations for Fe and
non-spin-polorized for Cr and Ni).}
\end{figure}

In Fig. \ref{fig:V2}, the quasibinary effective pair interactions,
$\widetilde{V}^{(2)-\alpha \beta}_p$, in Fe$_{75}$Cr$_{17}$Ni$_{08}$
and Fe$_{56}$Cr$_{21}$Ni$_{23}$ of Hamiltonian (\ref{eq:H_conf_qb})
are shown. They have been obtained for the high temperature
(~800--1000K) lattice constant of 3.615 \AA. Two sets of interactions
have been obtained in the DLM-LSF state at 800 K, while
one set of interactions, for Fe$_{56}$Cr$_{21}$Ni$_{23}$
alloy was calculated in the DLM calculations without LSF.
In this case, there is no magnetic moment on Ni and Cr atoms.
As one can see, the LSF produce quite substantial effect in
the case of Fe-Ni and Fe-Cr effective interactions, although
it is not that pronounced in the case of Ni-Cr interactions.
At the same time, the concentration dependence seems to
be quite moderate, within the range of the accuracy of the
SGPM calculations.

The dependence of the effective interactions on the lattice constant
is in fact quite strong (not shown in the figure). For instance,
the nearest-neighbor effective interactions obtained in the DLM
state for the theoretical 0 K lattice constant, 3.55 \AA, are
5.89, -0.62, and 10.05 mRy for Fe-Cr, Fe-Ni and Fe-Cr pairs,
respectively, which are quite different from those for
the high temperature lattice constant of 3.615 \AA: 3.96, 0.30,
and 8.24 mRy.
Obviously, the external and internal parameters cannot be
disregarded in finding theoretical atomic configuration of
austenitic steels at finite temperatures.

Effective interactions presented in Fig. \ref{fig:V2} look
qualitatively similar for all the pairs: the strongest interaction
of the ordering type at the first coordination shell following
the next strongest interaction with the opposite sign at
the second coordination shell. However, as we will see below, there
cannot be mutual ordering of all the pairs due to the fact that
the pairs with the strongest ordering energies, specifically NiCr
at the first coordination shell, will be ordering first. 

Let us note that the strong ordering interaction of Ni and Cr atoms
at the first coordination shell is almost entirely due to the screened
Coulomb interaction, which is of an order of 8--9 mRy, while the
one-electron contribution is only about 0.2--0.5 mRy. In the case
of Fe-Ni and Fe-Cr pairs, the nearest neighbor screened Coulomb interaction
is relatively small, about 1 and 2 mRy, respectively.
This is an expected result, since the screened Coulomb interactions
are proportional to the charge transfer between the corresponding alloy
components, which is in its turn determined mostly by their mutual size
difference. The latter is obviously the largest for Ni and Cr, while
it is relatively small for Fe-Cr and Fe-Ni pairs.

The size difference of alloy components leads to the appearance of
local lattice relaxations, which can be accounted for in the configurational
Hamiltonian by the so-called strain-induced interactions. Unfortunately,
it is impossible to calculate accurately these interactions in ternary Fe-Cr-Ni
alloys, especially at high temperatures in the paramagnetic state with LSF.
Therefore in this
work, we use a simple qualitative model. First of all, we disregard
the strain induced interactions for Fe-Ni and Fe-Cr pairs, which should
be small anyway. As for the Ni-Cr strain-induced interactions, we take
them from Ref. \onlinecite{rahaman14} where they have been obtained for
Ni-Cr binary alloys. Although this is a quite rough approximation, it is
used here to test a qualitative effect of such interactions onto the
ordering in austenitic steels.

Multisite interactions are relatively small in these alloys. The
strongest 3-site interaction is of the 114-type (for the nomenclature 
of the multisite interactions see Ref. \onlinecite{rahaman14}),
i.e. for the cluster of the three nearest-neighbors on the line in the
closed packed [110] direction (there are 8 such interactions in general
for a given 3-site cluster, which correspond to different combinations
of alloy indexes). It is especially strong for the CrCrCr[Ni]
configuration: 3.11 and 2.86 mRy in Fe$_{75}$Cr$_{17}$Ni$_{08}$
and Fe$_{56}$Cr$_{21}$Ni$_{23}$, respectively. This specific interaction
corresponds actually to the same type of interaction in Ni-Cr binary
alloy where it is in fact also the strongest 3-site interaction,\cite{rahaman14}
although the value of the interaction is quite reduced in 
ternary Fe-Cr-Ni alloys compared to that in binary Ni-Cr alloys.

The strongest 4-site interactions are for the tetrahedron of 
the nearest neighbors and for the four nearest neighbor sites 
along the closed packed [110] direction,
which are also the strongest interactions in the Ni-Cr
system.\cite{rahaman14} In ternary alloys, however, the quisibinary Fe-Ni
4-site interactions  (FeFeFeFe[Ni]) are comparable with the corresponding
quasibinary Ni-Cr interactions (CrCrCrCr[Ni]), while the interactions
with mixed alloy component indexes are almost an order of
magnitude smaller.\cite{V-4}

In order to check how the SGPM works, for this particular system,
the ordering energy of the Fe$_2$NiCr-L1$_2$m ordered structure 
in the DLM state (without LSF) have been calculated for the lattice
constant of 3.615 \AA ~from the EMTO total energies and from SGPM interactions
for Fe$_{50}$Ni$_{25}$Cr$_{25}$ random alloy. In the direct calculations
the ordering energy is $-$1 mRy/atom, while it is about $-$2 mRy/atom
from the pair, 3-site and 4-site SGPM interactions. The agreement is
reasonable, considering the smallness of the ordering energy, and the
fact that magnetic moment in the ordered state, $~1.0$ $\mu_{\rm B}$, is
different from that in the random alloy, $~1.6$ $\mu_{\rm B}$. Such a
small ordering energy ($~$160 K) also means that this structure can
hardly be formed at 650 K as predicted in Ref. \onlinecite{wrobel15}.

\begin{figure}
\includegraphics[width=0.9\linewidth]{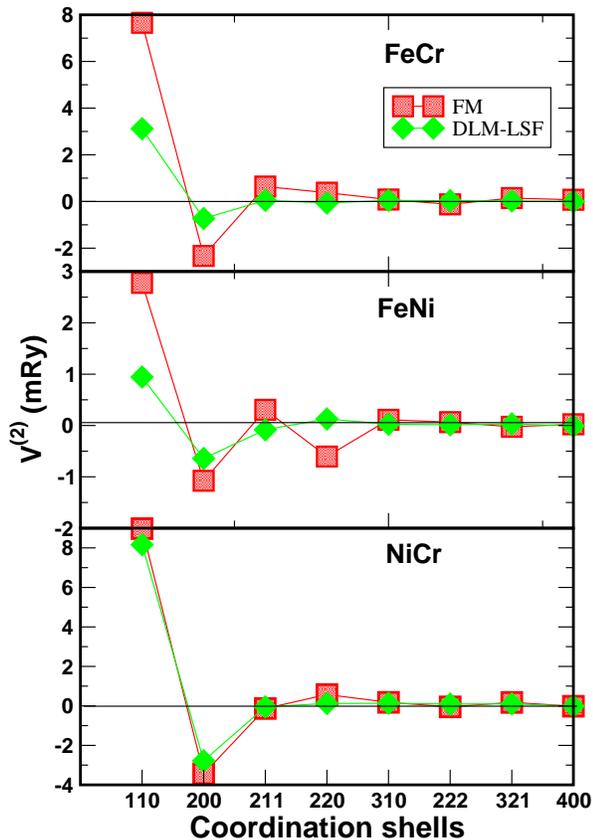}
\caption{\label{fig:V2fm} (Color online) 
Effective pair interactions in Fe$_{56}$Cr$_{21}$Ni$_{23}$ obtained
in the DLM-LSF state at 800 K and in the ferromagnetic 
(FM) state. 
}
\end{figure}

One of the reasons why the ordering strength is greatly
exaggerated in Ref. \onlinecite{wrobel15} is the fact that its
authors are using concentration independent cluster expansion for
0 K enthalpies of formation obtained in the ferromagnetic
state (with ferromagnetic alignment of Fe and Ni magnetic moments and 
antiferromagnetic one of Cr and Fe(Ni)) in the high temperature Monte
Carlo simulations.
To show the effect of the magnetic state upon the effective interactions,
we compare in Fig. \ref{fig:V2fm} the effective pair interactions for
random Fe$_{56}$Cr$_{21}$Ni$_{23}$ alloy obtained in the DLM-LSF
state at 800 K and in the ferromagnetic (FM) state (again, the magnetic
moment of Cr is antiferromagnetically aligned with those of Fe and Ni).

Although the Ni-Cr effective pair interactions are approximately
the same in both states, it is obviously not the case of
the Fe-Cr and Fe-Ni effective pair interactions: the strongest
nearest neighbor interaction for both pairs is more than twice
as large in the FM state as in the DLM-LSF state. Such a strong
dependence of the effective interactions on the magnetic 
state means that the corresponding modeling of austenitic
steels at finite temperatures should be done in the 
relevant to this temperature magnetic state.

Let us note that the effect of the magnetic state upon chemical
interactions in fcc Fe-Ni alloys have been already studied in Ref.
\onlinecite{ruban07_inv,ekholm10}. In particular, as has been demonstrated
in Ref. \onlinecite{ruban07_inv}, the DLM state leads to a significant
drop of the strongest nearest neighbor interaction in Invar
Fe$_{65}$Ni$_{35}$ alloy, which makes, in the end, the ordering
of this alloy impossible at temperatures relevant to its preparation.
It is interesting that the interactions obtained in Ref. 
\onlinecite{ruban07_inv} for the Invar binary alloy are very close
to those for Fe$_{56}$Cr$_{21}$Ni$_{23}$  shown in Fig. \ref{fig:V2fm}. 

It has been also demonstrated\cite{ekholm10} that
there is a substantial reduction of the effective chemical
interactions even in the finite temperature FM state due to a
reduce magnetization, which is very similar to the case of bcc
Fe-Cr alloys.\cite{ruban08} It is obvious that such a coupling
between magnetic state and effective interactions should be properly
taken into consideration in the corresponding thermodynamic
modeling at high temperatures.

\section{Atomic ordering in austenitic steels}

Monte Carlo calculations of atomic alloy configuration in
Fe$_{75}$Cr$_{17}$Ni$_{08}$ and Fe$_{56}$Cr$_{21}$Ni$_{23}$
alloys have been done using a simulation box containing
12$\times$12$\times$12($\times$ 4) sites. In both cases, the first 21
effective pair interactions have been used with the contribution from
the Ni-Cr strain-induced interactions taken from
Ref. \onlinecite{rahaman14}. We have also used the four strongest 3-site
interactions of the 111, 112, 113 and 114-type (see
Ref. \onlinecite{rahaman14} for the nomenclature), and two 4-site
interactions, for the tetrahedron of the nearest neighbors and the
nearest neighbor sites along the closed packed [110] direction.

The calculated atomic SRO parameters at 800 K are presented in
Table \ref{tbl:a_sro}
together with the experimental data from Ref. \onlinecite{cenedese84}.
Although the agreement is only qualitative, one should take into
consideration the fact that the strain-induced interactions are
considered only quite approximately in this work. Besides, the
experimental values seem to be quite sensitive to the model used
in the analysis of the diffuse scattering intensities. 

Nevertheless, the picture of atomic ordering is quite clear and consistent:
the strongest ordering is between Ni and Cr nearest and next nearest neighbors.
The calculated type of ordering is (100), which is the same as in the experiment,
but it is different from the (1$\frac{1}{2}$0) type in the ordered
Ni$_2$Cr phase. However, as has been shown in Ref. \onlinecite{rahaman14},
the effective cluster interactions and ordering are sensitive to the alloy
composition in fcc Ni-Cr alloys, and the (100)-type ordering is consistent
with the results for equiatomic Ni-Cr alloys.

The next in the strength of ordering are the pairs of Fe and Cr atoms at the
first coordination shell. Although the type of ordering in theoretical
calculations, (100) is not consistent with the experimental one, (1$\frac{1}{2}$0),
one can again notice that the type of ordering in the experiment depends on
the fitting model.\cite{cenedese84} We have also neglected the Fe-Cr
strain-induced interactions, which may produce a certain effect.

And finally, Fe and Ni atoms repel each other at the first coordination shell
exhibiting there a kind of "phase separation" behavior, although the corresponding
effective pair interaction is of an ordering type. As has been already
mentioned, the Fe-Ni effective pair interactions in binary Invar alloy,
Fe$_{65}$Ni$_{35}$,\cite{ruban07_inv} are very close to those obtained in
this work for austenite and shown in Fig. \ref{fig:V2}, where the slight ordering
tendency is observed. The difference between the atomic ordering of Fe and Ni
atoms in Invar alloy and ternary Fe$_{56}$Cr$_{21}$Ni$_{23}$ alloy is thus
entirely due to compositional restrictions in the latter case, which do not allow
atoms of all the types to establish their best local environment.

\begin{table}[htbp]
\caption{Warren-Cowley short-range-order parameters,
$\alpha_{lmn}^{\alpha \beta}$, determined in Monte Carlo simulations
for Fe$_{56}$Cr$_{21}$Ni$_{23}$ Ni at 800~K. The experimental
parameters are taken from Ref. \onlinecite{cenedese84}.
}
\label{tbl:a_sro}
\begin{ruledtabular}
\begin{tabular}{crrrrrr}
      & \multicolumn{2}{c}{$\alpha_{lmn}^{\rm FeCr}$} &
        \multicolumn{2}{c}{$\alpha_{lmn}^{\rm FeNi}$} &
        \multicolumn{2}{c}{$\alpha_{lmn}^{\rm NiCr}$} \\ 
\cline{2-7} \           
$lmn$  & Exp.      & Calc.    & Exp.      &    Calc. &   Exp. & Calc.     \\
\hline
 110   & $-$0.009  & $-$0.029 &    0.017  &    0.032 &  $-$0.113 & $-$0.235  \\
 200   &    0.043  &    0.032 & $-$0.002  & $-$0.012 &     0.148 &    0.261  \\
 211   &    0.029  &    0.003 & $-$0.002  &	   0.013 &     0.012 & $-$0.023  \\
 220   & $-$0.017  &    0.021 &    0.001  & $-$0.013 &     0.007 &    0.060  \\
 310   & $-$0.024  & $-$0.007 &    0.003  &	   0.004 &  $-$0.003 & $-$0.033  \\
 222   & $-$0.033  &    0.006 &    0.003  &	$-$0.004 &     0.012 &    0.021  \\
 321   &    0.005  & $-$0.004 &    0.002  &	$-$0.001 &  $-$0.004 & $-$0.003  \\
 400   & $-$0.029  &    0.008 & $-$0.002  & $-$0.001 &     0.026 &    0.019  \\
\end{tabular}
\end{ruledtabular}
\end{table}

It should be noted that the experimental data for the SRO parameters
in Fe$_{82-x}$Cr$_{18}$Ni$_{x}$
($x=$ 15, 20, and 25 wt.\%) alloys obtained by Braude {\it et al.}\cite{braude86}
for  temperatures $~$1400 K, seem to be quite strange. For instance,
the Warren-Cowley SRO parameter for Ni-Cr pairs is about $-$0.3 for
the first (110) and second (200) coordination shells, while it is about
~0.15 for the fourth (220) coordination shell. It is hard to imagine that
kind of ordering behavior and it also contradicts experimental data by
Cenedese {\it et al.}.\cite{cenedese84} The only common point is the
fact that the ordering tendency between pairs of alloy components
decreases in the order: Ni-Cr, Fe-Cr, and Fe-Ni.

\begin{figure}
\includegraphics[width=0.9\linewidth]{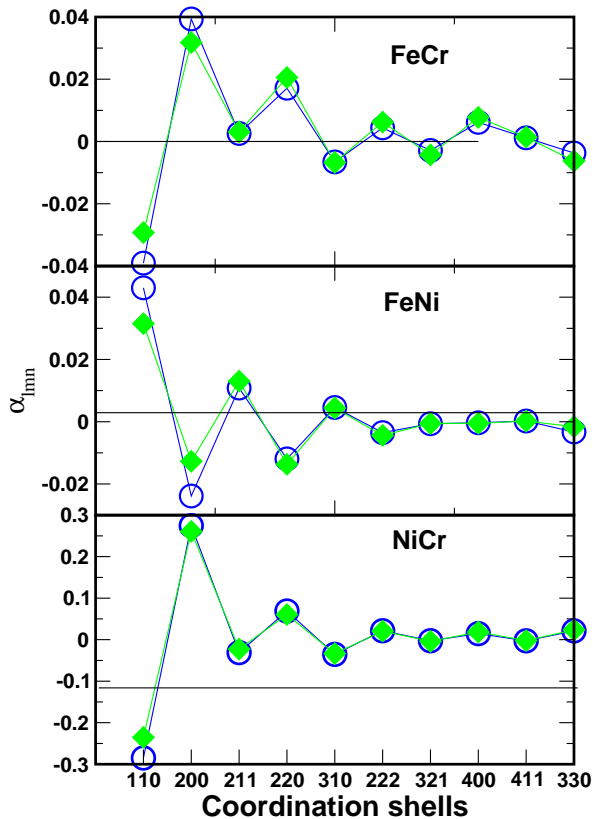}
\caption{\label{fig:a_sro} (Color online) 
Warren-Cowley short-range order parameters in 
Fe$_{75}$Cr$_{17}$Ni$_{08}$ (open circles) and
Fe$_{56}$Cr$_{21}$Ni$_{23}$ (filled diamonds) at 800 K
obtained in Monte Carlo simulations.
}
\end{figure}

The closeness of effective interactions and compositions for 
Fe$_{56}$Cr$_{21}$Ni$_{23}$ and Fe$_{75}$Cr$_{17}$Ni$_{08}$ alloys
means that the atomic SRO should be also very similar. In
Fig. \ref{fig:a_sro}, we show the atomic SRO in these 
alloys at 800 K. One can see that this is indeed the case:
there is quite strong ordering of Ni-Cr atoms. In both 
cases the values of the SRO parameters are quite large 
(in absolute value), which means that these alloys should
not be that far from an order-disorder phase transition.

\begin{figure}
\includegraphics[width=0.9\linewidth]{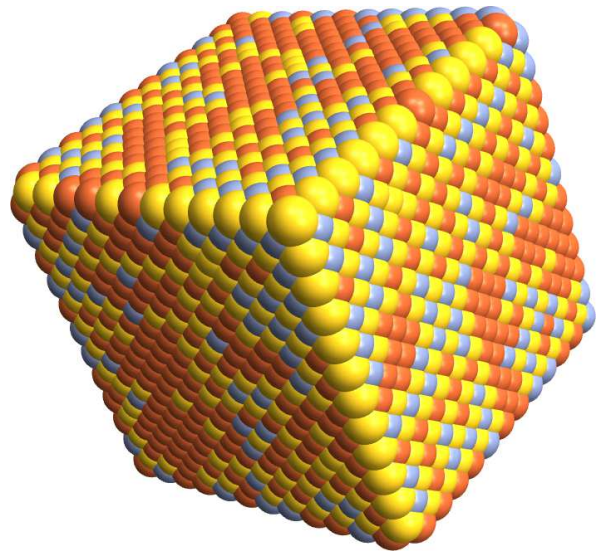}
\caption{\label{fig:ord_str} (Color online) 
Atomic structure of Fe$_{56}$Cr$_{21}$Ni$_{23}$ at 400 K
obtained in Monte Carlo simulations. Type of atoms are
colorcoded in the following way: Cr is yellow, Fe is red,
and Ni is grayblue. 
}
\end{figure}

Indeed, upon decreasing temperature the phase transition
into the (Fe,Ni)$_3$Cr-L1$_2$-like structure is observed at
about 520 K for the Fe$_{56}$Cr$_{21}$Ni$_{23}$ alloy
composition and at 480 K for the Fe$_{75}$Cr$_{17}$Ni$_{08}$
alloy composition, although most probably these transition
temperatures are overestimated since the theoretical atomic
SRO is stronger than experimental one. The ordered structure
is shown in Fig. \ref{fig:ord_str}
for the case of Fe$_{56}$Cr$_{21}$Ni$_{23}$ alloy (a similar
structure is observed in the Monte Carlo simulations
for Fe$_{75}$Cr$_{17}$Ni$_{08}$).

As one can see, Cr atoms occupy a simple cubic positions
of the fcc lattice, while Ni and Fe atoms occupy the remaining
(face-centred) positions forming a peculiar mutual ordering.
If the temperature in Monte Carlo simulations is lowed further,
there will another phase transition at about 170 K leading to a 
mixture of two ordered phases: Fe$_3$Ni$_3$Cr$_2$ and some not
so clear ordering paten of the remaining Fe and Cr atoms. However,
this last transition is hardly relevant to the reality, where 
low temperature magnetic state can change substantially
the interactions, and the atomic diffusion is absent on the usual
practical human time scale.

\section{Summary}

An accurate {\it ab initio}-based description of the finite-temperature
properties of austenitic stainless steels require taking the thermally
induced LSF into consideration. In this work, the simplest 
model has been used, which is based on a classical consideration of the
spin-fluctuation energy, which is a rough statical single-site
mean-field approximation. It is demonstrated, however, that it allows
one to bring such properties as bulk modulus in close agreement
with experimental data.

The simulations of atomic configuration of austenitic steels done
for two compositions, Fe$_{56}$Cr$_{21}$Ni$_{23}$ and
Fe$_{75}$Cr$_{17}$Ni$_{08}$, show that there is substantial degree
of the atomic SRO at high temperatures. These results are in
reasonable agreement with the existing experimental
data.\cite{cenedese84} The theoretical calculations by
Wrobel {\it et al.}\cite{wrobel15} predict much stronger ordering 
tendency due to the use of the low-temperature magnetic 
ground state in the cluster expansion. In this paper, we show that
Fe-Cr and Fe-Ni effective pair interactions at the first
coordination shell are approximately twice as large as those
in the DLM(-LSF) state.

In order to establish a truly quantitative picture of the atomic
ordering and the properties of austenitic steels at finite temperature,
further investigation is needed into the finite temperature magnetism
of these alloys and its use in the atomistic scale modeling of their
structural and vibrational properties.


\appendix
\section{SGPM interactions in the EMTO-CPA method}
\label{sgpm}

The one-electron contribution to the GPM interactions within the
EMTO-CPA Green's function formalism can be obtained from the 
one-electron energy, which in this case is (for details of the
EMTO-CPA parameters see \cite{vitos07})

\begin{equation} \label{E_1_el}
E_{one-el} = \frac{1}{2 \pi i} \oint z \langle G(z)\rangle dz ,
\end{equation} 
where $\langle G(z)\rangle$ is the CPA average Green's function of
alloy (for Brave lattice to simplify notations):

\begin{equation} \label{G}
\langle G(z)\rangle = \int_{\rm BZ} d {\mathbf k}
\sum_{LL'}\tilde{g}_{L';L}({\mathbf k},z)
\dot{S}_{L; L'}({\mathbf k},z)  + s.s.c.
\end{equation}
Here, only multisite term is
explicitly shown, while single-site contribution (s.s.c) is 
omitted since it does not contribute to the intersite chemical
and magnetic interactions. Summations are running over
sublattices ($ij$), indexes of angular momentum ($L,L'$) and
integration is done over k-points of the Brillouin zone.
$\dot{S}_{LL'}({\mathbf k},z)$
is the energy derivative of the slope matrix,
$S_{LL'}({\mathbf k},z$), which depend on energy, $z$,
and k-point, and 
$\tilde{g}_{LL'}({\mathbf k},z)$ is the k-point resolved CPA path
operator:
\begin{equation}
\tilde{g}_{LL'}({\mathbf k},z) =
\frac{1}{S_{LL'}({\mathbf k},z) - \tilde{D}_L(z)} ,
\end{equation}
where $\tilde{D}_L(z)$ is the coherent potential function of the
EMTO method obtained self-consistently from the following CPA
set of equations:

\begin{eqnarray} \label{CPA_eq}
&&\widetilde{g}_{LL'}(z) =
\int d {\mathbf k} \tilde{g}_{LL'}({\mathbf k},z) \equiv  \widetilde{g} \\
&&g^{\alpha} = \widetilde{g} + \widetilde{g} \left[D^{\alpha} - \tilde{D}\right]
g^{\alpha} \\
&&\widetilde{g} = \sum_{\alpha} c^{\alpha} g^{\alpha}
\end{eqnarray}
In the last two equations, we have omitted angular momentum index
and energy dependence of $\widetilde{g}$ and $D$.

The chemical and magnetic exchange interactions then can be found
using the force theorem \cite{FT} either for chemical fluctuations at
some particular site (relative to the CPA effective
medium)\cite{ducastelle76,gonis87,turchi88,drchal92,singh93,ruban04}
or by introducing small displacement of the direction of the spin
relative to unperturbed spin orientation at this
site\cite{oguchi83,licht84,licht87} and then finding
the change of the one-electron energy (\ref{E_1_el}) by expanding the
multisite part of the Green's function in (\ref{G}).

The resulting expressions are similar to those in the KKR
or KKR-ASA methods.\cite{ruban04} For instance, the GPM quasibinary
effective pair interactions in a multicomponent alloy are

\begin{equation} \label{eq:V^2}
\widetilde{V}^{(2)-\alpha \beta-1}_p = -\frac{1}{\pi} \Im \int^{E_F}
{ \rm Tr}  \left[ \Delta t^{\alpha \beta} \widetilde{g}_{ij}
\Delta t^{\alpha \beta} \widetilde{g}_{ji} \right] dE ,
\end{equation}
where $\Delta t^{\alpha \beta} = t^{\alpha} - t^{\beta}$, 
and $t^{\alpha}$ has the meaning of the single-site scattering
t-matrixes, which in the EMTO method are

\begin{equation}
t^{\alpha} = \left[ 1 + \widetilde{g}(\tilde{D} - D^{\alpha})  \right]^{-1}
(\tilde{D} -  D^{\alpha}) ,
\end{equation}
which actually satisfy the CPA equation:

\begin{equation}
\sum_{\alpha} c^{\alpha} t^{\alpha} = 0.
\end{equation}

Another quantity entering (\ref{eq:V^2}) is the CPA scattering
path operator between $i$ and $j$ sites, which belong to the
coordination shell $p$:

\begin{equation} \label{eq:g_ij}
\widetilde{g}_{ji}(z) = \int_{\rm BZ} 
\tilde{g}({\mathbf k},z) e^{i {\mathbf k}({\mathbf R}_i - {\mathbf R}_j)}
d {\mathbf k} ,
\end{equation}
where angular momentum indexes are omitted.

In general, the interaction of order $n$ of Hamiltonian (\ref{eq:H_conf})
is defined as

\begin{equation}
V^{(n)- \alpha \beta \ldots \gamma [\delta] }_{f} = 
-\frac{1}{\pi} \Im \int^{E_F} { \rm Tr}
\left[ t^{\alpha} \widetilde{g}_{ij} t^{\beta} \widetilde{g}_{jk} 
\ldots \widetilde{g}_{lk} t^{\gamma} \right] dE ,
\end{equation}
where $i$, $j$, $\ldots$, and $k$ are the sites of the cluster $f$.
As a matter of fact, this is only one specific contribution to this
interaction, and in order to get the total interaction for $n>3$,
one should sum over all possible paths connecting sites of the cluster.

This formalism can be easily generalized to the case of paramagnetic
alloys described by the DLM, or alloys with partial non-zero
magnetization within partial DLM (PDLM) approach. In this case,
the expressions for the effective interactions remain the same,
but $t^{\alpha}$ entering the corresponding formulas are modified.
In particular, if alloy component A is in the PDLM state
with magnetization $m < 1$, it is presented as an
alloy with spin-up and spin-down orientation,
A$^{\uparrow}_{x}$A$^{\downarrow}_{1-x}$, where $m = 2x -1$
(assuming that $x \geq 0.5$, and $x = 0.5$ corresponds to the
DLM paramagnetic state). It can be shown that corresponding 
magnetic averaging (like it is done, for instance, in Ref.
\onlinecite{ruban04,rahaman11}) results in this case in
$t^{A} = x t^{A^{\uparrow}} + (1-x) t^{A^{\downarrow}}$ for
every spin component.

Finally, in order to get the screened pair effective interaction
at the coordination shell $p$, one should add the corresponding
screening contribution, $V^{scr-\alpha \beta}_p$,
as it is defined in (\ref{eq:V_scr}), so that

\begin{equation}
\widetilde{V}^{(2)-\alpha \beta}_p =
\widetilde{V}^{(2)-\alpha \beta-1}_p + V^{scr-\alpha \beta}_p .
\end{equation}

\section{Magnetic exchange interactions in the EMTO-CPA method}

Magnetic exchange interactions, $J_p$ of Heisenberg Hamiltonian

\begin{equation} \label{eq:J_xc}
H^H = -\sum_p \sum_{ij \in p}J_p {\mathbf e}_i {\mathbf e}_j ,
\end{equation}
where ${\mathbf e}_i$ is the spin variable at site $i$, can
be derived using the so-called magnetic force theorem\cite{licht84,licht87}
and has the following form for elementary solid closely resembling
Eq. (\ref{eq:V^2}):

\begin{equation} \label{eq:_xc}
J_p = -\frac{1}{\pi} \Im \int^{E_F}
{ \rm Tr}  \left[ \Delta g_{ij}^{\uparrow}
\Delta g_{ji}^{\downarrow} \right] dE ,
\end{equation}
where $\Delta = D^{\uparrow} - D^{\downarrow}$,
and $g_{ij}^{\uparrow(\downarrow)}$ are the intersite path operator
for spin-up (and spin-down) states as they determined in (\ref{eq:g_ij}).

In the case of a random alloy, magnetic exchange interactions between
$\alpha$ and $\beta$ alloy components, $J_p^{\alpha \beta}$, is determined
as\cite{licht87}

\begin{equation} \label{eq:_xc}
J_p^{\alpha \beta} = -\frac{1}{\pi} \Im \int^{E_F}
{ \rm Tr}  \left[ \Delta^{\alpha}  \bar{g}_{ij}^{\alpha \beta \uparrow}
\Delta^{\beta} \bar{g}_{ji}^{\beta \alpha\downarrow} \right] dE ,
\end{equation}
where $\Delta^{\alpha} = D^{\alpha \uparrow} - D^{\alpha\downarrow}$
and

\begin{equation}
\bar{g}_{ij}^{\alpha \beta} =
\left[ 1 + \widetilde{g}(\tilde{D} - D^{\alpha})  \right]^{-1}
\widetilde{g}_{ij} \\ \nonumber
\left[ 1 + (\tilde{D} - D^{\beta})\widetilde{g}  \right]^{-1} .
\end{equation}
Here, it is assumed that alloy is homogeneous and thus there
is no site-dependence of the potential parameters (although
the generalization to inhomogeneous alloys is straightforward).

The DLM state is a special one since magnetic exchange interactions
in the DLM state are equal to the corresponding GPM interactions
$8 J_p = - \widetilde{V}^{(2)}_p$,\cite{licht87,shallcross05}
which can be proved analytically. On the other hand, it follows
from the comparison of the Ising and Heisenberg Hamiltonians and
the fact that the last one is reduced to the Ising one for collinear
magnetic configurations. This also means that GPM provides an
easy way to calculated higher order magnetic
interactions.\cite{ruban04,shallcross05}

\acknowledgments

AVR acknowledges the support of the Swedish Research Council
(VR project 2015-05538), the European Research Council grant,
the VINNEX center Hero-m, financed by the  Swedish Governmental
Agency for Innovation Systems (VINNOVA), Swedish industry, and
the Royal Institute of Technology (KTH). Calculations have been done
using NSC (Link\"oping) and PDC (Stockholm) resources provided
by the Swedish National Infrastructure for Computing (SNIC).
The support by the Austrian Federal Government
(in particular from Bundesministerium fŸr Verkehr, Innovation und
Technologie and Bundesministerium fŸr Wirtschaft, Familie und Jugend)
represented by …sterreichische Forschungsfšrderungsgesellschaft mbH
and the Styrian and the Tyrolean Provincial Government, represented
by Steirische Wirtschaftsfšrderungsgesellschaft mbH and Standortagentur
Tirol, within the framework of the COMET Funding Programme is
also gratefully acknowledged.


\begin{thebibliography}{99}
\bibitem{code} The Lyngby version of the EMTO code properly takes
into consideration electrostatics in random alloys in contrast to
other existing versions. It is distributed by the author of the
paper.
\bibitem{V_quasib} In its turn, the quasibinary pair interactions 
are 
$\widetilde{V}^{(2)-\alpha \delta}_p  = V^{(2)-\alpha \alpha [\delta]}_p$
and $\widetilde{V}^{(2)-\alpha \beta}_p  = V^{(2)-\alpha \alpha [\delta]}_p
+ V^{(2)-\beta \beta [\delta]}_p - 2 V^{(2)-\alpha \beta [\delta]}_p$ .
\bibitem{V-4} For example, the 4-site interactions for the tetrahedron
of the nearest neighbors are 1.54, -0.16, -0.09, -0.19, and 0.91 mRy
for FeFeFeFe, CrFeFeFe, CrCrFeFe, CrCrCrFe, and CrCrCrCr alloy indexes,
respectively.
\end{thebibliography}
\end{document}